\documentclass[superscriptaddress,showpacs,amssymb,twocolumn,aps,footinbib]{revtex4}

\usepackage{graphicx,dcolumn,bm,amsmath,color}

\newcommand{\eq}[1]{Eq.~(\ref{#1})}
\newcommand{\fig}[1]{Fig.~\ref{#1}}

\newcommand{\olcite}[1]{Ref.~\onlinecite{#1}}
\newcommand{\olcites}[1]{Refs.~\onlinecite{#1}}

\newcommand{\eea}{ \end{eqnarray} }
\newcommand{\bea}{ \begin{eqnarray} }

\newcommand{\eeq}{ \end{equation} }
\newcommand{\beq}{ \begin{equation} }

\newcommand{\bhua}{ \hat{\bf u} _{1} }
\newcommand{\bhub}{ \hat{\bf u} _{2} }

\newcommand{\bhu}{ \hat{\bf u} }

\newcommand{\bra}{ {\bf r}_{1}  }
\newcommand{\brb}{ {\bf r}_{2}  }
\newcommand{\br}{ {\bf r} }
\newcommand{\bs}{{\bf s }}
\newcommand{\bq}{{\bf q }}
\newcommand{\bk}{{\bf k }}

\newcommand{\bn}{ \hat{\bf n} }
\newcommand{\bx}{ \hat{\bf x} }
\newcommand{\by}{ \hat{\bf y} }
\newcommand{\bz}{ \hat{\bf z} }

\newcommand{\bv}{ {\hat {\bf v} } }
\newcommand{\bw}{\hat { {\bf w}} }

\newcommand{\kbt}{k_{\rm B}T}
\newcommand{\om}{\Omega}
\newcommand{\oma}{\Omega_{1}}
\newcommand{\omb}{\Omega_{2}}

\begin{document}

\title{Generalized Onsager theory for strongly anisometric  patchy colloids}

\author{H. H. Wensink}
\email{wensink@lps.u-psud.fr}
\affiliation{Laboratoire de Physique des Solides, CNRS UMR 8502,  Universit\'{e} Paris-Sud, 91405 Orsay Cedex, France}
\author{E. Trizac}
\affiliation{Laboratoire de Physique Th\'eorique et Mod\`eles Statistiques, CNRS UMR 8626, Universit\'{e} Paris-Sud, 91405 Orsay Cedex, France}

\date{\today}

\begin{abstract}
The implications of soft `patchy'  interactions on the orientational disorder-order transition of strongly elongated colloidal rods  and flat disks is 
studied within a simple Onsager-van der Waals density functional theory. The theory provides a generic framework for studying the liquid crystal phase 
behaviour of highly anisometric cylindrical colloids, which carry a distinct geometrical pattern of repulsive or attractive soft interactions localised 
on the particle surface. In this paper, we apply our theory to the case of  charged rods and disks for which the local electrostatic interactions can 
be described by a screened-Coulomb  potential.  We consider infinitely thin rod-like cylinders with a uniform line charge and infinitely thin discotic 
cylinders  with several distinctly different surface charge patterns.   Irrespective of the backbone shape, the isotropic-nematic phase diagrams of 
charged colloids feature a generic destabilization of nematic order, a dramatic narrowing of the biphasic density region and a 
reentrant phenomenon upon reducing the electrostatic screening.  At higher particle density  the electrostatic repulsion leads to a complete suppression of nematic order 
in favour of spatially inhomogeneous liquid crystals.  
 
\end{abstract}

\maketitle

\section{Introduction}

Many colloidal dispersions, such as natural clays, and (bio-)macromolecular systems consist of rod- or disk-shaped mesogens whose intrinsic ability to form liquid crystalline 
order gives rise to unique rheological and optical properties \cite{gennes-prost}. Despite their abundance in nature, the statistical mechanics of fluids containing 
anisometric particles in general (and oblate ones in particular) has received far less attention than that of their spherical counterparts. The possibility of a first-order 
disorder-order transition from an isotropic to a nematic phase was first established theoretically by Onsager \cite{Onsager} in the late 1940s. Although originally devised for 
rod-like particles in solution, his theory also makes qualitative predictions for plate-like particles based on the central idea that orientation-dependent harshly repulsive interactions 
alone are responsible for stabilizing nematic order. Subsequent numerical studies have fully established the phase diagram  of hard prolate 
\cite{taylor_prl1989,Bolhuisintracing,mcgrother,annebohle_prl1996,graftmv,graflowen_jpcm1999} and oblate hard cylinders \cite{Veerman,wensink_mp2009,marechal_cuetos2011}. 
Owing to the simplicity of the interaction potential, hard-body systems constitute an essential benchmark for the study of liquid crystals and their phase stability.   
Temperature becomes merely an irrelevant scaling factor in the free energy and the phase behaviour is fully determined by the volume fraction occupied by the particles and the aspect ratio.  
At high volume fraction, additional entropy-driven disorder-order transitions  occur where a nematic fluid transforms into positionally ordered phases \cite{Frenkel88}. 
Depending on the cylinder aspect ratio the system may develop a smectic phase, characterized by a one-dimensional periodic modulation along the nematic director, or a 
columnar phase consisting of columns with a liquid internal structure self-assembled into a two-dimensional crystal lattice. Similar to nematic order, the formation 
of smectic, columnar or fully crystalline structures is based entirely on entropic grounds \cite{frenkellc}; the loss of configurational entropy associated with (partial) 
crystalline arrangement  is more than offset by a simultaneous increase in translational entropy, that is, the average free space each particle can explore becomes larger in the ordered phase.

In most practical cases, however, particle interactions are never truly hard and additional enthalpic contributions impinge on 
the free energy of the system. Long-ranged 
interactions usually originate from the presence of surface charges leading to  electrostatic repulsions between colloids \cite{verwey1955theory,hansenlowen2000}, 
or from traces of 
other colloidal components such as non-adsorbing polymers, which act as  depletion agents and give rise to effective attractive interactions \cite{Vrijdepletie,lekkerkerker2011colloids}.  
Other site-specific interactions may originate from  hydrogen-bonding \cite{takashi_2000} or end-functionalized polymers such as DNA grafted onto the colloid surface \cite{geerts_eiser2010}.   
Depending on their nature (repulsive or attractive), interaction range, and topological arrangement on the particle surface, these site-specific directional interactions  
may greatly affect the self-assembly properties of anisometric particles  \cite{mcgrother_sear1997,demichele_bellini2012,avendano_2009}. 
In this context it is also worth mentioning recent progress  in the fabrication of anisometric colloids with `patchy' interactions \cite{glotzer_spheroid2013,glotzer_plates2013} 
where the interplay between patchiness and the anisometric backbone shape offers a rich and intriguing repertoire of novel structures \cite{glotzer_solomon2007}. 

These recent developments suggest the need for a comprehensive theory for lyotropic systems  which explicitly accounts for these patchy interactions. 
The aim of the present paper is to set up such a theory  by combining the classic Onsager theory for slender hard bodies, with a mean-field van der Waals treatment for the 
additional long-ranged interactions \cite{cotter1977,cotter1979,gelbartbaron,vargachiral2006,francomelgar2008}. Most molecular-field type theories developed to date focus 
on rod-like mesogens with dispersion interactions represented by an orientation-dependent potential with some radially symmetric spatial variation, akin to a Maier-Saupe 
form \cite{maiersaupe1,luckhurst1977}. Here,  we shall lay out the framework for the more general case of slender rod and disk-shaped cylinders 
carrying site interactions with arbitrary integrable form and spatial arrangement. 
By exploiting  the simple second-virial structure of the Onsager reference free energy, we show  that these soft patchy interactions, 
on the mean-field level, give rise to a 
non-trivial orientation-dependent van der Waals (or molecular field) term which strongly affects the disorder-order transition in the fluid state. 

We illustrate its practical use by focusing on isotropic-to-nematic and nematic-to-smectic or columnar phase transitions in systems of charged prolate and discotic colloid 
in the salt-dominated regime, a subject of considerable research interest given that natural clays consist of strongly charged colloids.  The majority of clays are composed 
of sheet-like minerals colloids \cite{vanolphen1963,davidson-overview2005} but rod-shaped mineral colloids may display similar properties 
\cite{lemaire_davidson2004,roorda_vanblaaderen2004,zhang_duijneveldt2009}.  It is still largely unclear how the interplay between particle shape and electrostatics controls the structure 
and dynamics of clay systems. The fundamental understanding is further complicated by the fact that both the magnitude and sign of the local charge density may vary significantly along 
the particle surface. For instance, under certain chemical conditions laponite platelets \cite{Mourchid1995} adopt opposite face and rim-charges and the intrinsic patchiness of the 
electrostatic interactions may lead to unusual liquid behaviour  \cite{ruzicka2011}.  Incorporating these patchy  interactions into a state-of-the-art statistical physical machinery to 
extract structural information remains a daunting task. Headway can be made by using computer simulation where  a number of  coarse-grained  models for non-isometric charged colloids have 
been studied over the past decade \cite{ganzenmueller2010,trizac_jpcm2002,morales2012,delhorme_jpcl2012,degraaf2012,jabbari2013}.  

With the present theory, we aim to set a first step towards linking microscopic patchiness of soft interactions to liquid crystal stability for strongly anisometric colloids. 
We apply the generalized Onsager theory to the case of  charged cylinders interacting through an effective Yukawa potential and demonstrate  a generic destabilization and 
non-monotonic narrowing of the biphasic gap upon reducing the electrostatic screening.  The influence of the geometric pattern of the charge patches can be incorporated 
explicitly by means of a form factor, as shown for disklike colloids. The present calculations, however, merely serve an illustrative purpose and the main goal is to open up 
viable routes to studying more complicated surface charge architectures of clay nano sheets \cite{delhorme_sm2012,odriozola2004,kutter2000,leger2002} or anisotropic 
Janus particles \cite{walther_mueller2008,cheng_janus2008}. Moreover, the theory can be further refined by using effective parameters, pertaining to the backbone shape, 
charge density, screening constant etcetera, in order to enable more quantitative predictions for highly charged anisometric colloids.
 
Although the Onsager treatment is strictly limited to low to moderate density, it offers possibilities to assess the stability of high-density liquid crystal phases on 
the level of a simple bifurcation analysis  \cite{wensink2007}.  We show that it is possible to extend the generalized-Onsager form into a full density functional form  
(e.g. using judicious parametric form for the one-body density) is possible. This holds promise for incorporating soft interactions into more sophisticated hard-body density 
functionals such as those based on fundamental measure theory \cite{hansengoos_mecke2009,hansengoos2011,cuesta_jcp1997}, weighted-density approximations \cite{graflowen_jpcm1999}, 
renormalized Onsager theories \cite{parsons1979,francomelgar2008},  or cell-theory \cite{taylor_prl1989,graftmv}.  The use of reliable non-local reference free energy functionals is 
expected give a more quantitative account of patchy rods or disks with broken translational symmetry induced by a high particle density, geometric confinement \cite{rosenfeld_schmidt_pre1997} 
or surfaces \cite{bier_jcp2006}. The generalized Onsager theory bears some resemblance to other interaction-site models such PRISM/RISM theories \cite{costa_mp2005, harnau_mp2008} which have 
been invoked to study the thermodynamic properties of isotropic plate fluids, but have not yet proven capable of treating liquid crystal phases at higher particle densities. The effect of 
attractive interparticle forces on the bulk phase behaviour of ionic liquid crystals has been scrutinized in \olcite{kondrat_jcp2010} using a mean-field theory of  the Gay-Berne potential 
for ellipsoidal mesogens.

The remainder of this paper is structured as follows. In Section II, we outline the mean-field Onsager theory for soft patchy cylinders with vanishing thickness,
while the specific features of the screened Coulomb potential are addressed in section III.
The theory will then be applied in Section IV to study the isotropic-nematic phase diagram of charged rod- and disklike cylinders in the strong screening regime. 
Possible ways to include spatially inhomogeneous  liquid crystals into the generalized 
Onsager treatment are highlighted in Section V. Finally, some concluding remarks are formulated in Section VI.

\section{Mean-field Onsager theory for soft patchy potentials}

Let us consider a system of $N$ infinitely thin colloidal cylindrical disks or rods with length $L$ and diameter $D$ at positions $\{ \br^{N} \}$ and orientations 
$\{ \om^{N} \}$ in a 3D volume $V$ at temperature $T$. We assume the particle shape to be maximally anisotropic so that the aspect ratio $L/D \rightarrow \infty$ 
(infinitely elongated rods) and $L/D \downarrow 0 $ (infinitely flat disks). In the fluid state, the particle density $\rho = N/V$ is homogeneous throughout space.  
Following Onsager's classical theory \cite{Onsager} we may write the Helmholtz free energy as follows: 
\beq
\frac{\beta F}{N}  \sim \ln {\mathcal V} \rho  + \langle  \ln 4 \pi f(\om) \rangle - \frac{\rho}{2} \left \langle \left \langle \int_{V} d \br \Phi ( \br ; \oma , \omb) \right \rangle \right \rangle,
\eeq
with $\beta^{-1} = \kbt$ in terms of Boltzmann's constant $k_{B}$ and ${\mathcal V}$ the total thermal volume  of a cylinder,
defined from the cube of the de Broglie wavelength, dressed by contributions from the rotational 
momenta since the kinetic energy not only consists of translational terms. The value of ${\mathcal V}$ will prove immaterial in the subsequent analysis.
The brackets denote  orientation averages $\langle  \cdot \rangle = \int d \om f(\om ) (\cdot ) $ and $\langle  \langle \cdot \rangle \rangle = \iint d \oma d \omb f(\oma ) f(\omb) (\cdot ) $ 
in terms of the orientational distribution function (ODF) $f(\om)$ which expresses the probability for a cylinder to adopt a solid angle $\om$  on the 2D unit sphere. The shape of the ODF allows us to 
distinguish between isotropic order,  where $f = 1/4\pi $,  and nematic order where $f$ is some peaked function. Particle interactions are incorporated 
on the second-virial level via a spatial integral over the Mayer function:
\beq
\label{mayer}
\Phi( \br ; \oma , \omb) = e^{ -\beta U ( \br ; \oma , \omb)}  - 1,
\eeq
which depends on the pair potential $U$ between two cylinders with centre-of-mass distance $ \br = \bra  - \brb $. In our model we shall assume each particle to consist 
of a cylindrical hard core (HC) with diameter $D$ and height $L$ supplemented with a soft interaction potential $U_{s}$ describing (effective) long-ranged interaction 
with neighboring particles. These soft interactions can either be repulsive or attractive and may originate from effective interparticle forces between the colloids under 
the influence of depletion effects \cite{lekkerkerker2011colloids}, polymers end-grafted onto the colloid surface \cite{likos_pr2001} or electrostatics \cite{verwey1955theory}. 
The corresponding potential is unlikely to be a simple radially symmetric function but rather emerges from a particular spatial arrangement of interaction sites located on the 
cylinder surface. In the latter case the soft potential is given by a summation over site-site interactions which are assumed to have a radially symmetric 
form $u(r)$ \footnote{This form represents a simplified subset of more general orientation-dependent segment potentials of the form $u(\br ; \oma^{(l)}, \omb^{(m)})$, 
such as for e.g. segment dipoles, where $u$ depends on the orientation $\om_{1}^{(l)}$ of site vector $l$ with respect to the molecular frame of particle 1. }
\beq
\label{segpot}
 U_{s}(  \br  ; \oma, \omb)  =  \sum_{l,m}  u( |  \br + \bs_{l}(\oma ) - \bs_{m}(\omb ) | ),
\eeq
where $\bs_{i}$ denotes the distance vector between site $l$ located on the surface of cylinder 1 and the centre-of-mass $\br_{1}$. 
The total pair potential thus reads:  
\beq
 U (  \br  ; \oma, \omb) =
  \begin{cases}
  \infty &  \text{if hard cores overlap} \\
  U_{s}(  \br  ; \oma, \omb)  & \text{otherwise}  \label{hcpot}.
  \end{cases}
\eeq
For hard cylinders ($U_{s}=0$), the spatial integral over the Mayer function yields the excluded volume between two cylinders at fixed orientations. In the limit of maximal cylinder anisotropy, one obtains \cite{Onsager}:
\beq
v_{\text{excl}} (\gamma) = -\int_{V} d  \br \Phi_{HC} ( \br ; \oma , \omb) =   v_{0} | \sin \gamma |,  \label{vexcluded}
\eeq
with $v_{0} = 2L^{2} D $ for needles ($ L/D \rightarrow \infty $) and $v_{0} = \pi D^{3}/2$ for disks ($L/D \downarrow 0$).  $\gamma(\oma , \omb)$ denotes the enclosed angle between 
the normal vectors of two cylinders. The total free energy of the fluid can be compactly written as:
\bea
\label{fbare}
\frac{\beta  F}{N} &  \sim & \ln {\mathcal V} \rho  + \langle  \ln 4 \pi f(\om) \rangle +\frac{\rho}{2} \left \langle \left \langle v_{\text{excl}}(\gamma) \right \rangle \right \rangle \nonumber \\ 
&+& \frac{\rho}{2} \left \langle \left \langle \int _{\br \notin v_{\text{excl}}}d  \br \left ( 1 - e^{ - \beta U_{s}( \br ; \oma , \omb )}  \right ) \right \rangle \right \rangle. 
\eea 
The spatial integral in the final term runs over the space complementary to the finite excluded volume manifold formed by the hard cores of two cylinders at fixed orientations. 
The last term can be interpreted as an {\em effective} excluded volume but a direct calculation of this quantity poses some serious technical difficulties \cite{eggen2009}.  
A more tractable expression can be obtained by adopting a mean-field form which can be obtained  by taking the limit $\beta U_{s} \ll 1 $ in the second-virial term.  \eq{fbare} can then be recast into a form resembling a generalised van der Waals free energy:
\bea
\label{vdw}
\frac{\beta  F}{N} &  \sim & \ln {\mathcal V} \rho  + \langle  \ln 4 \pi f(\om) \rangle + \frac{\rho}{2} \left \langle \left \langle v_{\text{excl}}(\gamma) \right \rangle \right \rangle \nonumber \\ 
&+& \frac{\beta \rho}{2} \left ( a_{0}   -  \left \langle \left \langle a_{1}(\oma , \omb ) \right \rangle \right \rangle \right ),
\eea 
where the contributions $a_{0}$ and $a_{1}$ can be identified as van der Waals constants emerging from  spatial averages of the soft potential. The non-trivial one, $a_{1}$, 
runs over the excluded volume manifold of the cylinders:
\beq
\label{a1real}
 a_{1}(\oma , \omb) =  \int _{\br \in v_{\text{excl}}} d  \br    U_{s} ( \br ; \oma , \omb ),  
\eeq
whereas $a_{0}$  represents an integration over the entire spatial volume $V$:
\bea
\label{viso1}
a_{0} &= &  \int_V d \br  U_{s}( \br ; \oma , \omb)  \nonumber \\
& = &    \sum_{l,m} \int _{V} d  \br u( |  \br + \bs_{l}(\oma ) - \bs_{m}( \omb) | ). 
\eea 
Introducing a linear coordinate transformation $ {\bf y} \rightarrow   \br + \bs_{l}(\oma) - \bs_{m}(\omb)  $ (with Jacobian unity) yields a trivial constant:
\beq
\label{viso2}
a_{0}  =    \sum_{l,m} \int _{V} d  {\bf y} u( |  {\bf y} | ) 
  = 4 \pi  \int_{0}^{\infty} dr r^{2} u(r)  =  \text{cst},
\eeq 
{\em independent} of the mutual cylinder orientation. In arriving at \eq{viso2}, we have tacitly assumed that the spatial integral over the soft part of the pair potential 
is bounded. For this to be true,  the 3D Fourier transform (FT) of the site potential must exist:
 \beq
 \label{ftex}
 \hat{u}(q) = 4 \pi \int_{0}^{\infty} dr r^{2} \frac{\sin qr}{qr} u(r).
 \eeq
This requires that the potential be less singular than $1/r^{3}$ in the limit $r \downarrow 0$  and decay sufficiently fast to zero  as $r \rightarrow \infty$. 
Steep repulsive potentials such as those associated with dipolar interactions, $u \sim 1/r^{3}$, or van der Waals dispersion forces, $u \sim -1/r^{6}$ \cite{israelachvili}, do not qualify 
and our treatment is aimed at potentials such as the screened-Coulomb (Yukawa) potential \cite{verwey1955theory} or various bounded potentials such as  
Gaussian \cite{berne_jcp1972,likos_sm2006}, square-well \cite{bolhuis_jcp1997,delrio_jcp2005} or  linear  ramp potentials which routinely arise from free-volume type theories 
for depletion interactions \cite{lekkerkerker2011colloids} or as effective potentials for end-grafted polymers \cite{likos_pr2001}.  
We remark that the free energy \eq{vdw} represents a hybrid between the second-virial approach, which is valid at low particle densities, and the mean-field approximation, 
accurate at high particle density. For charged cylinders it will be shown that the theory represent a simplified alternative to a more formal variational hard-core 
Poisson-Boltzmann theory for anisometric hard cores developed in \olcite{lue_fpe2006} and a coarse-grained density functional theory for Yukawa fluids in \olcite{hatlo_lue_jcp2012}. 

We shall now proceed with analysing the non-trivial van-der-Waals contribution \eq{a1real}. In view of the existing FT  it is expedient to recast the spatial integral in \eq{vdw} 
in reciprocal space. The analysis is further facilitated by using the linear transform introduced right after \eq{viso1}. After some rearranging the angle-dependent van der 
Waals term, \eq{a1real} can be factorized  in Fourier space in the following way:
\bea
\label{a1}
a_{1} (\oma, \omb)   &=&  \frac{1}{(2 \pi)^{3}}
   \int d \bq  \hat{u}(q) W(\bq ; \oma )W (-\bq ; \omb) \nonumber \\
   && \times   \hat{v}_{\text{excl}} (\bq ; \oma , \omb),
\eea
in terms of the FT of the excluded volume manifold of two cylinders (calculated in the Appendix):
\bea
\hat{v}_{\text{excl}}(\bq ; \oma, \omb) &=& \int_{\br \in v_{\text{excl}}} d \br e^{i \bq \cdot \br} \nonumber \\
&=& v_{0} |\sin \gamma |  {\mathcal F} (\bq ; \oma, \omb),
\eea
where the expressions  for ${\mathcal F}$  are given explicitly in the Appendix. The contribution $W$  pertains to a FT of the spatial resolution of the interaction sites  
according to: 
\bea
\label{ww}
W(\bq ; \om _{\alpha}) &=& \sum_{l} e^{i \bq \cdot \bs_{l}(\om _{\alpha})}, \hspace{0.3cm} \alpha =1,2 
\eea
which may be interpreted as a  {\em form  factor} reflecting the internal structure of the interaction sites on the particle surface. The simplest case,  a point segment 
located at the centre-of-mass thus corresponds to $\bs _{1}= \bs_{2} =  {\bf 0 }$ so that $W=1$. More complicated  configurations shall be considered in the following Section.

Next, the equilibrium form of the ODF is obtained by a formal minimization of  \eq{vdw}  : 
\beq
\frac{\delta}{\delta f} \left (  \frac{\beta F}{N} - \lambda \langle 1 \rangle \right ) = 0, 
\eeq
where the Lagrange parameter $\lambda$ ensures the ODF to be normalised on the unit sphere. The associated self-consistency equation for the ODF reads:
\beq
\label{odfnum}
f(\oma ) = {\mathcal Z}^{-1} \exp \left [  -\rho \left \langle (v_{\text{excl}}(\oma, \omb) - \beta a_{1}(\oma, \omb) ) \right \rangle _{2}  \right ],
\eeq
with normalisation constant ${\mathcal Z} = \langle \exp [\cdot ] \rangle_{1} $ and $\langle \cdot \rangle _{j}  = \int d \Omega_{j} f(\Omega_{j}) (\cdot )$ a one-body average over the angular set $\Omega_{j}$ at fixed $\Omega_{i\neq j}$.
It is easy to see that the isotropic solution $f= \text{cst}$,  i.e., all orientations  
being equally probable, is a trivial solution of the stationarity condition. Beyond a critical particle density non-trivial nematic solutions will appear which can be obtained 
by numerically solving \eq{odfnum} \cite{herzfeldgrid}.  Once the equilibrium ODF is established for a given density  phase equilibria between isotropic and nematic states can 
be investigated by equating the pressure $P$ and chemical potential $\mu$ in both states. These are obtained by standard thermodynamic derivatives of the free energy \eq{vdw}:
\bea
\label{presmu}
\beta P &=& \rho + \frac{\rho^2}{2} \left \langle \left \langle  v_{\text{excl}}(\oma , \omb) + \beta a_{0} - \beta a_{1}(\oma , \omb ) \right \rangle \right \rangle  \nonumber \\
\beta \mu &=&  \ln \rho {\mathcal V} +  \langle  \ln 4 \pi f(\om ) \rangle \nonumber \\
&& + \rho \left \langle \left \langle  v_{\text{excl}}(\oma , \omb) + \beta a_{0} - \beta a_{1}(\oma , \omb ) \right \rangle \right \rangle.
\eea
The  thermodynamic properties of the isotropic-nematic transition can be calculated by numerically solving these coexistence equations in combination with \eq {odfnum}, the stationarity 
condition for the ODF.  Collective orientation order of cylinders with orientation unit vector $\bhu$ order can be probed by introducing a common nematic director $\bn$ and defining nematic
 order parameters such as:
\beq
S_{n} = \langle {\mathcal P}_{n}(\bhu \cdot \bn) \rangle, 
\eeq
where $\mathcal P_{n}$ represents a  $n$th-order Legendre polynomial (e.g. ${\mathcal P_{2}}(x)=(3x^{2} - 1)/2$). Odd contributions of $S_{n}$ are strictly zero for non-polar phases and $S_{2}$ 
is routinely used to discriminate isotropic order ($S_{2}=0$) from uniaxial nematic order $S_{2} \neq 0$.

\section{Generalized screened-Coulomb potential for cylinders}

In this section, we shall analyze a simple model for charged anisotropic colloidal particles. Let us consider two disk-shaped macro-ions with total surface charge $Z$ in a 
electrolyte solution with  ionic strength determined by the counter ions and additional co- and counter ions due to added salt. Formally, the electrostatic potential 
around the charged surface of a macro-ion  in an ionic solution with a given ionic strength can be 
reasonably obtained from the non-linear Poisson-Boltzmann (PB) equation \cite{verwey1955theory}. 
This theory neglects any correlations between micro-ions and assumes the solvent  to be treated as a continuous medium with a given dielectric constant.  In the Debye-H\"{u}ckel approximation, valid if the 
electrostatic potential at the macro-ion surface is smaller than the thermal energy, the PB equation can be linearized  and the electrostatic interaction between two point
macro-ions with equal charge $\pm Ze$ in a dielectric solvent with relative
permittivity $\varepsilon_{r}$ is given by the screened-Coulomb or Yukawa form:
\beq
\label{yuk}
\beta u_{0}(r)= Z^{2}\lambda_{B}  \frac{e^{- \kappa r}}{r},
\eeq
with $\varepsilon_{0}$ the dielectric permittivity  in vacuum, $r$
the distance between the macro-ions, $\lambda_{B} = \beta e^{2}/4\pi
\varepsilon_{0}\varepsilon_{r} $ the Bjerrum length
($\lambda_{B} = 0.7$ $nm$ for water at $T=298K$) and $\kappa^{-1}$ 
the Debye screening length which measures the extent of the electric double layer.  In the limit of strong electrostatic screening,  the screening factor 
is proportional to $ \kappa = (8 \pi  \lambda_{B}  n_{0})^{1/2} $ with $n_0$ the concentration of added 1:1 electrolyte. In general, for highly charged colloids, non-linear effects of the 
PB equation can be accounted for by invoking a cell approximation \cite{ACGM84,deserno-holm2001} which assumes a fully crystalline structure where each particle is compartmentalized in Wigner-Seitz 
cells  or a so-called Jellium model \cite{trizac_levin_pre2004} where a tagged particle is exposed to a structureless background made up by its  neighboring particles.  Both methods allow 
for a solution of the full non-linear PB equation for an isolated colloidal with the effect of the surrounding charged particles  subsumed into a suitable boundary condition.  This procedure 
yields so-called {\em effective} values for the charge $Z_{\text{eff}} < Z$ and Debye screening constant $\kappa_{\text{eff}}$ which can be used to achieve accurate predictions for the 
thermodynamic properties (e.g. osmotic pressure) of fluids of highly charged spheres \cite{dobnikar_njp2006}.  We will briefly touch upon these effective parameters in paragraph C of 
this section. We reiterate that we focus here on the high-salt regime where use of the linearized form \eq{yuk} combined with effective electrostatic parameters is deemed appropriate. 
The low-screening regime requires a lot more care due the fact that the effective interaction becomes inherently dependent on the macroion density. As a consequence,  the free energy 
contains non-trivial volume terms which may have important implications for the fluid phase behaviour \cite{levin_trizac_jpcm2003,trizac_belloni_pre2007,zoetekouw_pre2006}.  

The FT of the Yukawa potential is given by a  simple Lorentzian: 
\beq
\label{ftyuk}
\hat{u}(q) =  Z^{2} \lambda_{B} \frac{4 \pi}{q^{2} + \kappa^{2}}.
\eeq
The spatial average over \eq{yuk} yields for $a_{0}$:
\beq
\beta a_{0} = 4 \pi Z^{2} \lambda_{B} \kappa^{-2}.
\eeq
We may generalize the screened-Coulomb potential for a cylindrical object by imposing that the total effective electrostatic potential be given by a sum over $n$ identical Yukawa sites 
located on the cylinder surface. As per \eq{segpot} the pair potential is given by: 
\beq
\beta U_{s}  =  \frac{Z^{2} \lambda_{B}}{n^2}  \sum_{i,j < n} \frac{\exp [ - \kappa | \br + \bs_{i}(\oma) - \bs_{j}(\omb) |}{  | \br + \bs_{i}(\oma) - \bs_{j}(\omb) |  }.
\eeq
We note that the generalized form is only adequate for non-isometric hard cores with a vanishing internal volume  \cite{trizac_jpcm2002}.  Such objects can be obtained, e.g., by taking  a cylindrical object with an extreme aspect ratio. 
In the next paragraphs, we shall analyze the expression above for infinitely elongated rods  ($D/L \downarrow 0$) and subsequently for  flat cylindrical disks ($L/D \downarrow 0 $).
  
  \subsection{Needle limit}
 
In case of infinitely slender charged rods, we assume a continuous distribution of sites located along the normal unit vector $\bhu $ running through the centre-of-mass of the cylinder. 
The result is a double integration along the one-dimensional contours of the rod pair. Defining a dimensionless contour parameter  $\ell_{i}$, so that $\bs (\om_\alpha ) = \ell_\alpha L \bhu _ \alpha $ 
the generalized screened-Coulomb potential between rodlike particles can be written as:
\beq
\label{elrod}
\beta U_{s} = Z^{2} \lambda_{B} \int_{-\frac{1}{2}}^{\frac{1}{2}}  d \ell_{1} \int _{-\frac{1}{2}}^{\frac{1}{2}} d \ell_{2} \frac{\exp [ - \kappa | \br + L( \ell_{1} \bhua - \ell_{2} \bhub ) |}{  | \br +  L ( \ell_{1} \bhua - \ell_{2} \bhub )|  }.
\eeq
A closed-form solution of the electrostatic rod potential was reported in \olcite{askari_jpcm2011} and a generalized DLVO form for rodlike macro-ions has been analyzed in  \olcites{chapot_jcp2004,chapot_jcis2005}. 
A tractable form for the electrostatic potential between infinitely stretched linear charges was used  by Onsager in his seminal paper \cite{Onsager,stroobants_mm1986}  based on a limiting form for  $\kappa L \rightarrow \infty$ \cite{fixman_skolnick1978}.  

Since our focus is on a simple van der Waals description for uniform fluids ($\rho = \text{cst}$), the compound form of  \eq{elrod}  naturally deconvolutes into a spherically 
symmetric kernel $\hat{u}(q)$ and form factor $W$ (cf.  \eq{a1}).  It is obvious that such a factorization becomes much more complicated in the  columnar, smectic or crystalline 
states where inhomogeneities in the density field are intricately coupled to the distance-variation of the electrostatic potential. This we shall see in more detail in Section V.   
We may specify the form factor by considering a linear array of interaction sites \cite{schneider_jpa1984}. In the continuum limit,  \eq{ww} becomes:
\bea
W_{\text{needle}}(\bq ; \om _{\alpha}) &=& \int_{-\frac{1}{2}}^{\frac{1}{2}}  d \ell_{\alpha} e^{-i L \bq \cdot \bhu_{\alpha}}  \nonumber \\ 
& = & j_{0} \left ( \frac{L}{2} \bq \cdot \bhu_{\alpha} \right ), \hspace{0.1cm} \alpha =1,2
\eea
with $j_{0}(x) = \sin{x}/x$  a spherical Bessel function. With this the orientation-dependent van der Waals constant \eq{a1} for rods is completely specified. 
The remaining 3D integration over reciprocal space must be carried out numerically for every orientation.     Note that  an evaluation in real space would confront us with 
a five-fold numerical integration since \eq{elrod} cannot be solved analytically.  To facilitate the integration over $q$-space, we adopt a particle-based frame $\{ \bhua, \bhub , \bv \} $ 
introduced in the Appendix. This allows  us to reexpress the dot products in terms of the angle $\gamma$ between the main axis of the rod pair via:
\bea
\label{dotr}
L\bq \cdot \bhua &=& q_{1} + q_{2} \cos \gamma  \nonumber \\
L\bq \cdot \bhub &=& q_1 \cos \gamma + q_2 \nonumber \\
D \bq \cdot \bv &=& q_3,
 \eea 
and $ \int d \bq  = (L^2 D)^{-1} | \sin \gamma | \prod_{i \leq 3} \int_{-\infty}^{\infty} dq_{i} $. The integration over $q_3$ can be carried out analytically and the remaining 
expression can be simplified by taking the leading order contribution in the needle limit $x=D/L \ll 1 $. The mean-field contribution 
$a_{1}(\gamma)$ for strongly elongated charged rods then reads in normalized form:
\bea
\frac{a_{1}(\gamma)}{a_{0}} &=& \frac{1}{4 \pi^{2} } (1 - e^{-\kappa D}) \sin^{2} \gamma  \int_{-\infty}^{\infty} dq_1  \int_{-\infty}^{\infty} dq_2 \nonumber \\ 
&& \times   j_{0}^{2} (L\bq \cdot \bhua )    j_{0}^{2} (L\bq \cdot \bhub ). \label{a1explicit}
\eea
Recalling that $a_{0} \propto \kappa^{-2} $ one can infer that $a_{1}$ vanishes in the limit of infinite screening $\kappa D \rightarrow \infty $  and diverges in the 
Coulomb limit ($ \kappa D \downarrow 0$) \cite{deutsch_jp1980} as one would intuitively expect. 

\subsection{Flat disk limit}

\begin{figure}
\begin{center}
\includegraphics[clip=,width= 0.9\columnwidth]{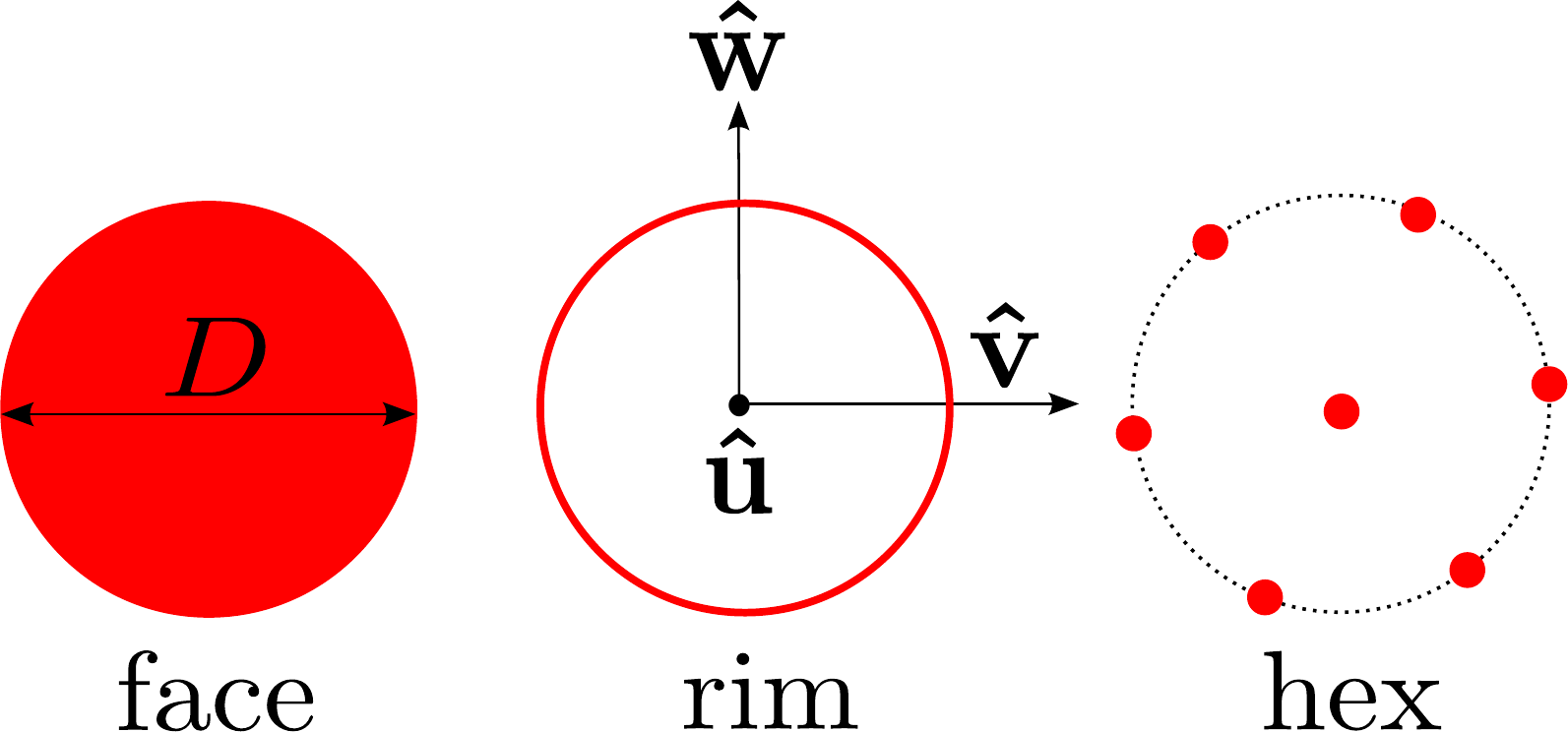}
\caption{ \label{fig1} Sketch of three possible surface charge patterns for infinitely thin disks. From left to right: uniform distribution over the circular face with 
diameter $D$ (``face"), one-dimensional distribution along the outermost circular contour (``rim") and a discrete hexagonal arrangement (``hex"). }
\end{center}
\end{figure}

We now turn to the case of infinitely thin disks. As for the needles, we assume a continuous charge distribution along circular surface of the disk,  which is most 
conveniently parameterized by invoking the particle-based coordinate frame (see Appendix) so that:
\beq
\label{splate}
\bs ( {\om_{\alpha}}) = \frac{D}{2} r _{\alpha} ( \bv \cos \xi_{\alpha} + \bw_{\alpha} \sin \xi_{\alpha} ), 
\eeq
where $ 0 \le r_{\alpha} \le 1$ and $0 \le \xi_{\alpha} \le 2 \pi $. The electrostatic potential between two flat disks at fixed orientations is represented by a four-fold integral:
\bea
\label{wplate}
\beta U_{s} &=& Z^{2} \lambda_{B} \frac{1}{\pi^2}\prod_{\alpha =1,2} \int_{0}^{1}  d r_{\alpha} r_{\alpha} \int _{0}^{2 \pi} d \xi_{\alpha} \nonumber \\
&& \times  \frac{\exp [ - \kappa | \br + \bs ( \oma ) - \bs ( \omb ) |}{  | \br + \bs ( \oma ) - \bs ( \omb )  |  }.
\eea
The form factor associated with a discotic arrangement of surface charges is given by the cosine transform of \eq{splate}: 
\bea
W_{\text{face}}(\bq ; \om _{\alpha}) &=& \frac{1}{\pi} \int_{0}^{1}  d r_{\alpha} r_{\alpha}  \int_{0}^{2 \pi} d \xi_{\alpha} \cos (\bq \cdot \bs(\om_{\alpha})) \nonumber \\
& = &  2   J_{1} ( \tilde{q}_{\alpha} ) / \tilde{q}_{\alpha},
\eea
with $J_{n}(x)$ a Bessel function of the first kind and $ \tilde{q}_{\alpha} =  \left [( \frac{D}{2} \bq \cdot \bw_{\alpha}  )^{2} + ( \frac{D}{2} \bq \cdot \bv  )^{2} \right ] ^{1/2}$. 
We may also consider the situation where the charges are distributed along the circular rim of the disk (\fig{fig1}). The corresponding form factor simply follows from  \eq{splate} and \eq{wplate} by 
setting $r_{\alpha} = 1$ and integrating over the remaining angular part:
\bea
W_{\text{rim}} (\bq ; \om _{\alpha}) &=& \frac{1}{2 \pi}  \int_{0}^{2 \pi} d \xi_{\alpha} \cos (   \frac{D}{2} ( \bv \cos \xi_{\alpha} + \bw_{\alpha} \sin \xi_{\alpha} )) \nonumber \\
& = &   J_{0} ( \tilde{q}_{\alpha} ). 
\eea
Alternatively, we may consider a discrete hexagonal arrangement of surface charges (see \fig{fig1}), in which case the form factor becomes: 
\bea
W_{\text{hex}}(\bq ; \om _{\alpha})  &=& \frac{1}{7} \left ( 1 + \cos(D \bq \cdot \bv) + 2  \cos (\frac{D}{2} \bq \cdot \bv ) \right . \nonumber \\ 
&& \left .  \times \cos (\frac{D \sqrt{3}}{2} \bq \cdot \bw_{\alpha}    ) + (\bv \leftrightarrow \bw_{\alpha}) \right ).
\eea
The last term ensures that the form factor remains invariant with respect to a rotation in the $\bv, \bw_{\alpha}$ plane so that $W_{\text{hex}}$ attains the same symmetry as the expressions 
for the ``face'' and ``rim'' patterns. The additional angular correlations naturally arise from the discrete nature of the hexagonal pattern. In view of the fluid phases considered here, they are deemed of negligible importance. 
We remark that all form factors  approach the radially  symmetric limit ($W=1$) in the macroscopic limit $q \downarrow 0$.  Similar to the needle case the FT  definition of the van der 
Waals contribution \eq{a1} reduces the dimensionality of the problem to a straightforward integration over 3D $q$-space whereas the real-space route would confront us with an intractable  
seven-fold integration. Analogous to \eq{dotr}, the integration over reciprocal space can be parameterized using the particle-based frame for disks (see Appendix):
$D\bq \cdot \bw_{1} = q_{1} + q_{2} \cos \gamma $, 
$ D\bq \cdot \bw_{2} = q_1 \cos \gamma + q_2 $, and
$ D\bq \cdot \bv = q_3 $ so that  $a_{1}$ depends only on enclosed angle $\gamma$ between the normal vectors of the disks as should be the case for apolar uniaxial cylinders. 

\begin{figure}
\begin{center}
\includegraphics[clip=,width= 0.8\columnwidth]{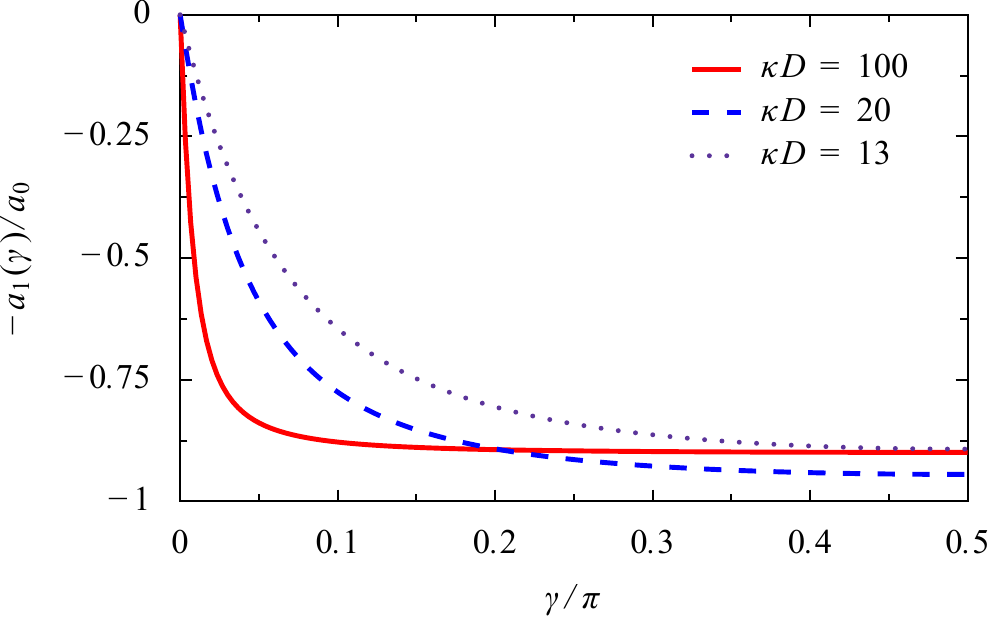}
\caption{ \label{fig2}   Aligning potential of mean force (in dimensionless units) for infinitely thin Yukawa disks with a homogeneous distribution of screened charges covering the circular surface of the disk. Shown are curves for different ionic strengths $\kappa D$.  
Near-parallel disk orientations  ($\gamma \sim 0$) are strongly disfavored.  }
\end{center}
\end{figure}

Judging from \eq{odfnum}, it is evident that $a_{1}$ can be identified with an aligning potential of mean force $V_{\text{mf}}(\om)$, reflecting the average potential  incurred by the soft 
potential of all the surrounding cylinders. This potential is inherently density-dependent and reads: 
\beq
V_{\text{mf}} ( \om) =  -\rho \int d \om^{\prime} a_{1}( \gamma ( \om, \om^{\prime} )) f( \om^{\prime} ),
\eeq
 where the ODF $f$ depends implicitly on $\rho$. Its angular dependence generally disfavours parallel orientations, as  illustrated in \fig{fig2} for the case of disks with a continuous 
 distribution of Yukawa sites (``face''). Similar monotonically decreasing functions are obtained for rods with the screening constant $\kappa D$ governing the typical range of the 
 potential. We reiterate that $a_1(\gamma)$ represents a distance-averaged orientational mean-field potential which is only applicable in the context of uniform isotropic or nematic fluids.   On the other hand,the spatially resolved electrostatic potential for charged disks bears an intricate coupling between the mutual orientation and centre-of-mass separation distance of the disk pair such that, 
 at least in the far-field limit, coaxial pair configurations are favored over planar  ones (see \eq{farfield} in Section III-C) \cite{trizac_jpcm2002,AgTB04}. At high particle density, 
 the interplay between near-field steric  and far-field electrostatic forces may drive the formation of liquid crystalline structures with unusual positional and orientational 
 microstructure  \cite{jabbari2013, morales2012}, thereby significantly affecting the stability of smectic and columnar order as we shall see in Section V.

\subsection{Second-virial coefficient for highly charged disks}

In this Section, we shall look at an alternative route towards incorporating electrostatic interactions into the Onsager density functional theory for the case of highly 
charged discotic colloids.  The objective is to make an estimate of the total second-virial coefficient of a charged disk. The orientational dependence  of this quantity 
gives us an idea of the effective shape (anisometry) of a charged discotic object and its propensity to form orientationally ordered phases at various screening conditions \cite{eggen2009}.    
To circumvent the computational burden  associated with a spatial integration of the Mayer function for segment potentials, we shall consider a tractable form for the electrostatic potential that can be obtained from non-linear (as well as linearized) Poisson-Boltzmann theory.  
In  
the {\em far-field} limit, the latter potential for anisotropic colloids can be recast into the following form \cite{trizac_jpcm2002,AgTB04,AlTe10}:
 \beq
 \label{farfield}
 U_{s}(\br ; \oma , \omb) = Z_{\text{eff}}^{2} \lambda_{B} \xi(\kappa D, \vartheta_{1}) \xi (\kappa D, \vartheta_{2}) \frac{ e^{-\kappa r}}{  r }.
 \eeq
The anisotropy function $\xi(\kappa D , \vartheta)$ depends on the screening parameter $\kappa$ and the angle $\vartheta $ between the centre-of-mass distance vector $\hat{\br}$ 
and disk normal $\bhu$ such that $\cos \vartheta = \hat{\br} \cdot \bhu$. Generally $\xi$ increases with  $\vartheta$ and reaches a maximum at $\vartheta = \pi/2$.   \eq{farfield} 
tells us that the orientation-dependence of the electrostatic potential is retained in the far-field limit,  and that stacked pair configurations are energetically favored over co-planar ones, irrespective of the centre-of-mass separation distance $r$.    

For highly charged colloidal disks, the strong coupling between the macro- and micro-ion 
charges leads to non-linear effects (such as counterion condensation) which can be quantified from the non-linear PB equation. The non-linearities can be taken into account by 
replacing the bare charge by an {\em effective} renormalized charge $Z_{\text{eff}}$. Its saturation value depends on the screening parameter and can be estimated as 
$Z_{\text{eff}}^{\text{sat}} \lambda_{B} /D \approx 0.5 \kappa D + 1.12$ \cite{AlTe10}.  An approximate form for the anisotropy function is given by \cite{trizac_jpcm2002}:
\beq
\xi (\kappa D , \vartheta) = 2 \frac{I_{1} \left (\frac{\kappa D}{2} \sin \vartheta \right )}{\frac{\kappa D}{2} \sin \vartheta},
\eeq
with $\sin \vartheta = (1 - (\hat{\br} \cdot \bhu)^{2})^{1/2}$ and $I_{1}(x)$ a modified Bessel function of the first kind.
Improved expressions can be found in \cite{AlTe10}.

Within Onsager's original second-virial approximation, the excess free energy is proportional to the second-virial coefficient $B_{2}$ embodied by the last two terms of \eq{fbare}. 
The excess free energy can thus be compactly written as:
\beq
\label{b2}
 \frac{\beta F_{ex}}{N} =  - \frac{\rho}{2} \left \langle \left \langle \beta_{1} (\oma , \omb) \right \rangle \right \rangle = \rho B_{2},
\eeq 
where the  cluster integral $\beta_{1}$ is given by  a spatial integral of the Mayer function \eq{mayer}. For the electrostatic part we need  to integrate over the space complementary 
to the excluded volume between two infinitely thin disks for which we may invoke the parameterization \eq{cuboid} proposed in the Appendix. The cluster integral then becomes:
\bea
\label{yukvir}
 \beta_{1} (\gamma) & =& - v_{\text{excl}}(\gamma) \nonumber \\ 
 &&  +  \left (  \int _{V} d \br  -  \prod_{i=1,3} \int_{-1}^{1} dt_{i}  J_{cc}  \right )   \Phi( t_{i} ; \gamma ),  
\eea  
with $J_{cc} = \frac{D^{3}}{8}| \sin \gamma | [ ( 1- t_{1}^{2} )^{1/2} + (1 - t_{2}^{2})^{1/2} ] $ the Jacobian associated with the transformation from the Cartesian lab frame to the 
particle frame and $\Phi$ the Mayer function given by \eq{mayer}. Since both volume integrals are defined within the latter frame, the orientation degrees of freedom naturally condense 
into a single angle $\gamma$ between the disk normals. Comparing \eq{yukvir} to the van der Waals form for patchy cylinders \eq{a1}  we see that both expressions  involve a 3D  
integration in Fourier or real space which can be numerically resolved without  difficulty.

\begin{figure*}
\begin{center}
\includegraphics[clip=,width= 1.6\columnwidth]{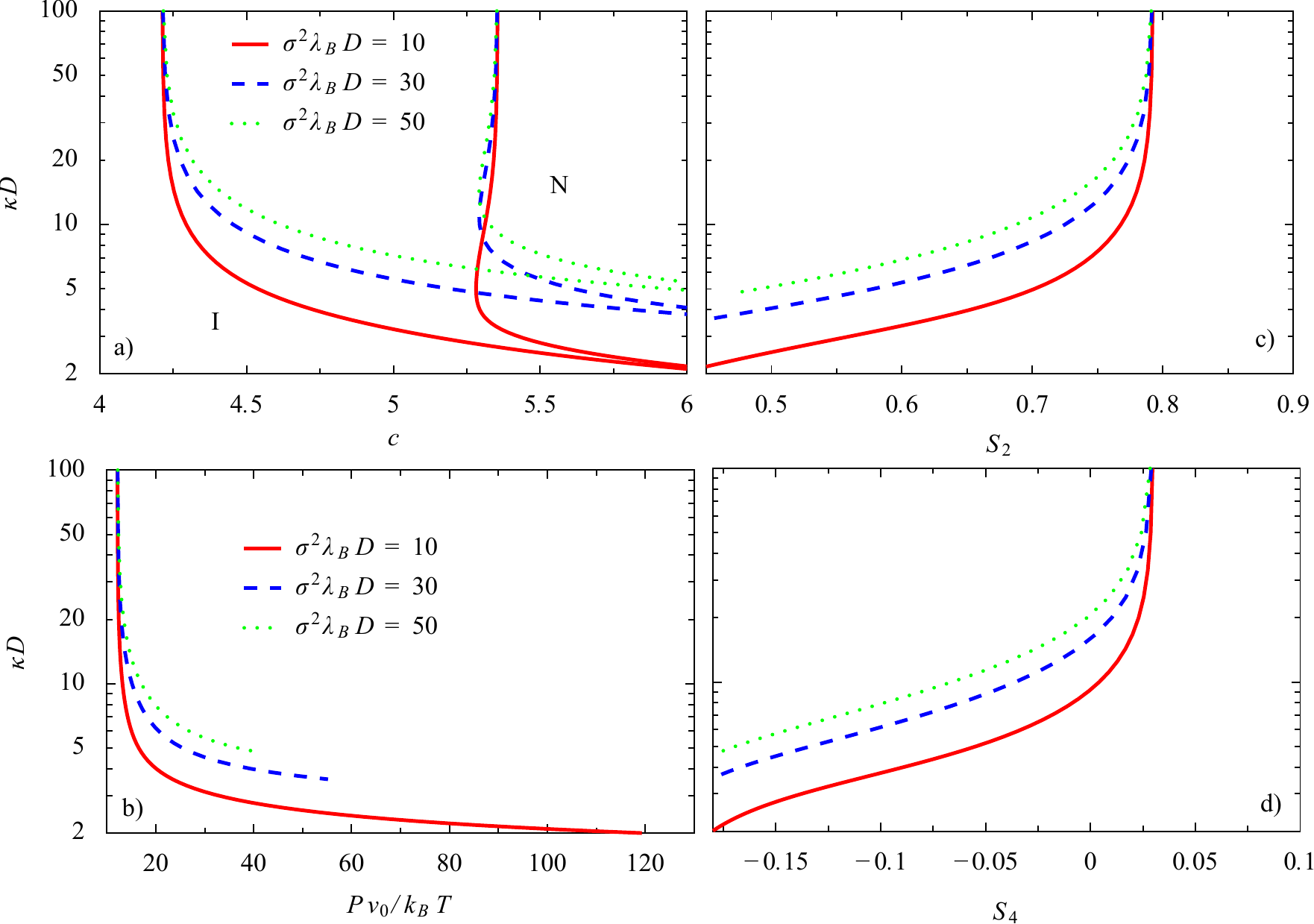}
\caption{ \label{fig3} (a) Isotropic-Nematic phase diagram of charged elongated colloidal rods ($L/D \rightarrow \infty $) for three different values of the 
Yukawa amplitude $ \sigma^{2}  \lambda_{B} D $. Plotted are (a) the coexistence densities $c=\rho L^{2}D$  versus ionic strength $ \kappa D $.  (b) Osmotic pressure $P$ at coexistence. 
(c-d) Orientational order parameters $S_{2}$ and $S_{4}$ quantifying the nematic  and cubatic  order of the nematic phase.}
\end{center}
\end{figure*}

\section{Results for the isotropic-nematic transition}

\begin{figure}
\begin{center}
\includegraphics[clip=,width= 0.8\columnwidth]{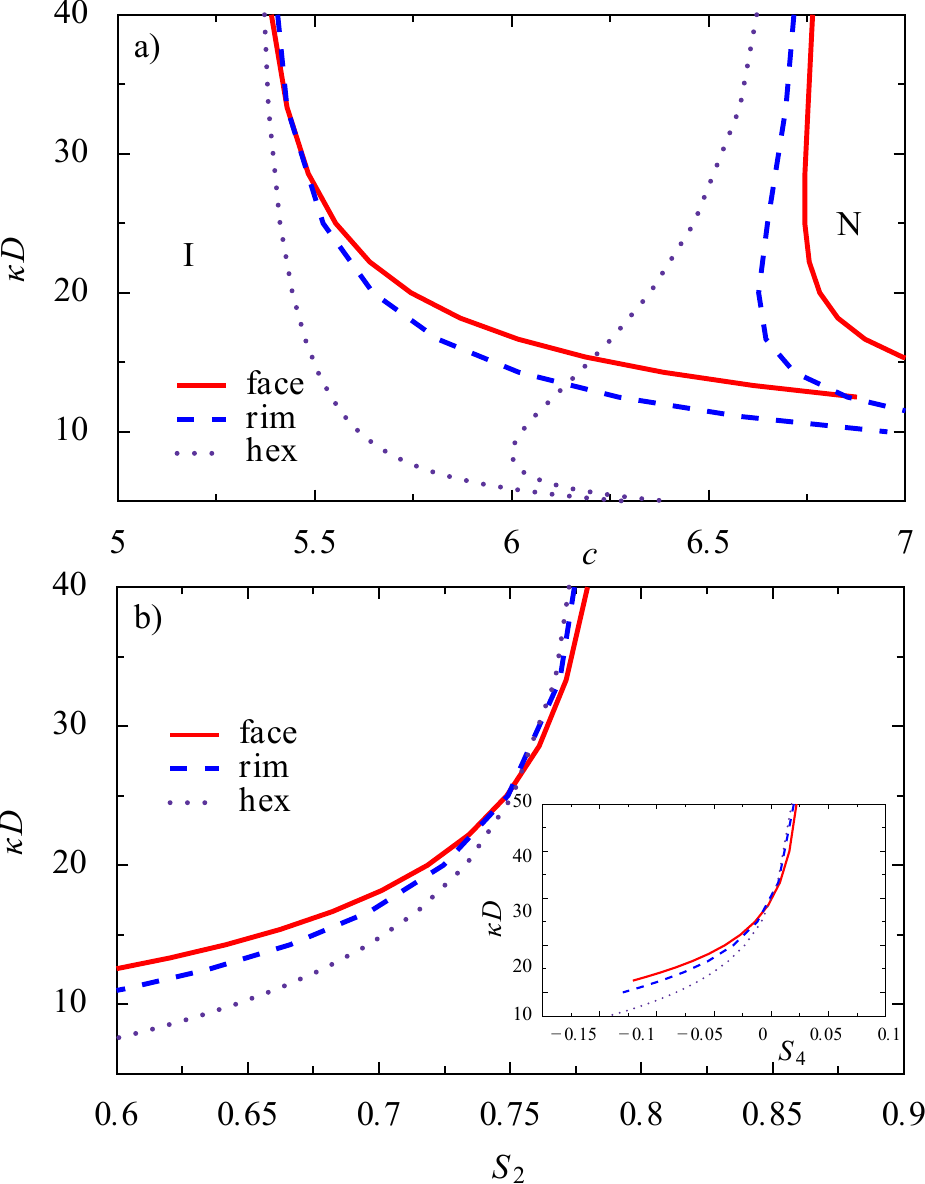}
\caption{ \label{fig4} (a) Isotropic-Nematic binodals for charged disks with charge distributions corresponding to the patterns indicated in \fig{fig1}. Plotted is the particle 
concentration $c=\rho D^{3}$ versus the ionic strength $\kappa D$. The Yukawa amplitude is $Z^{2} \lambda_{B}/D = 10$. (b) Orientational order parameters of the nematic phase at coexistence. }
\end{center}
\end{figure}

We now turn to the isotropic-nematic phase diagram for charged cylinders in the extreme aspect ratio limit.  Let us first concentrate on the case of infinitely 
elongated rodlike cylinders with  $L/D \rightarrow \infty$. The physical quantities of interest are the dimensionless concentration $c = \rho L^{2}D$, the charge $Z$, and the 
amplitude of the screened Coulomb potential. It is customary to define a linear charge density $\sigma$ indicating the number of elementary charges per unit length so that the total 
rod charge $Z = \sigma L$ leads to a dimensionless amplitude $ \sigma^{2} D \lambda_{B} $.
If we take a typical rod diameter of $D\sim 10 nm$ the Yukawa amplitudes in \fig{fig3}  corresponds to a linear charge density $\sigma$ of several elementary charges per $nm$. 
Furthermore we consider the case of excess salt so that the screening constant $\kappa$ does not depend on the colloid concentration.  The phase diagram in \fig{fig3} features a 
dramatic narrowing of the biphasic gap at lower ionic strength and a significant weakening of nematic order of the coexisting nematic phase.  
The negative sign of $S_{4}$ reflects an increased propensity for the rods to adopt perpendicular pair configurations at low screening so as to 
minimize the overlap of their electric double layers. This is a manifestation of the so-called ``electrostatic twist" for line charges which has been quantified in detail 
in  \olcite{stroobants_mm1986}.  The rapid variation of the binodal densities at low ionic strength reveals a marked re-entrant phase separation effect. A homogeneous isotropic 
sample at fixed particle density (say $c \sim 5.3$) undergoes a sequence of phase transformations upon increasing the ionic strength. First, the system exhibits isotropic-nematic phase coexistence with a weak density contrast.  
Second, the sample reverts to a homogeneous isotropic state before re-entering into a  phase-separation with a strong density difference between the coexisting phases.  
The isotropic-nematic 
transition disappears below a critical screening constant which is roughly independent of the line charge.  
The narrowing of the phase gap and upward shift of the transition density as the strength of the electrostatic interaction potential increases are both generic features of charged 
anisometric colloids,  consistent with predictions from previous mean-field theories for rods in the Coulomb limit  \cite{deutsch_jp1980}, and at finite screening  \cite{stroobants_mm1986}.

We remark that most colloidal systems consist of highly charged colloidal objects and that non-linear effects arising from the PB equation must be taken into account. As alluded to before, 
this can be done by retaining the linearized Debye-H\"{u}ckel form  and using a ``dressed" renormalized line charge $\sigma_{\text{eff}}$ which depends, in general, on the macro-ion density, 
shape and salt concentration. Despite the highly non-trivial  relation between these quantities, it is possible to derive simple analytical estimates for the saturation value, such that, for 
strongly elongated cylinders one can write $\sigma_{\text{eff}}^{\text{sat}} \lambda_{B} \sim   \kappa D $ \cite{trizac_aubouy_jpcm2003}.  However,  within the current scheme no stable nematic phase is found when simply replacing $\sigma$ by $\sigma_{\text{eff}}$ in case of  strong screening since $ (\sigma_{\text{eff}}^{\text{sat}})^{2}\lambda_{B}D \gg 1$ and the 
isotropic-nematic transition will be completely obstructed by the denematizing mean-field potential $a_{1}(\gamma)$.  Two remarks are in order.
First, the absence of a thermodynamically stable nematic phase could be a spurious result of the present theoretical set-up that may be remedied, at least in part,  by devising a more refined free energy by
carrying over part of the harshly repulsive near-field  electrostatic potential into the second virial coefficient e.g. by introducing an effective diameter $D_{\text{eff}}  > D$.  This opens up ways 
to designing optimized schemes that combine an effective particle shape with an appropriately rescaled aligning background potential capturing the far-field electrostatics at high particle density.  
These ideas have been pursued in detail in \olcites{lue_fpe2006,hatlo_lue_jcp2012} and \olcite{eggen2009} and shall not be further discussed here.      
   Second, the complete destabilization of spatially homogeneous orientational order may hint at the presence of more complicated types of nematic order in systems of highly charged rods. In particular, one could imagine the antagonistic effect of short-ranged aligning forces (due to the rod excluded-volume) and long-ranged dealigning ones (embodied in the mean-field potential $a_{1}$) to facilitate the formation of nematic phases with strongly inhomogenous, e.g.,  cubatic, biaxial or periodically modulated director fields \cite{sokolovska_prl2003}.
The occurrence of these complex textures remain to be confirmed by experiments or large scale computer simulations of charged rods in the strong coupling limit.

\begin{figure}
\begin{center}
\includegraphics[clip=,width= 0.8\columnwidth]{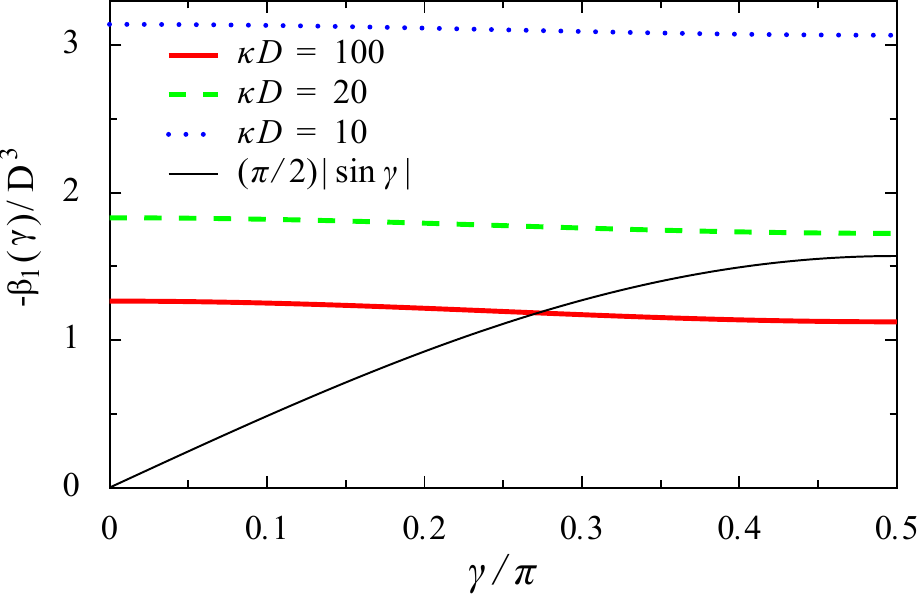}
\caption{ \label{fig5}  Effective excluded volume $-\beta_{1}(\gamma)$ between highly charged disks with diameter $D= 35 \lambda_{B}$ interacting via the orientation-dependent Yukawa 
potential \eq{farfield}.  The black solid line indicates the bare excluded volume of hard disks.   }
\end{center}
\end{figure}

Let us now turn to the case of charged disks. The isotropic-nematic phase diagram emerging from the Onsager-van der Waals theory for the various  charge patterns depicted in \fig{fig1} 
is shown in \fig{fig4}. Similar to the case of rods we observe a marked weakening of  nematic order and a narrowing of the biphasic gap.  The overall shape of the binodals does not depend 
too sensitively on the amplitude   provided that $Z^{2} \lambda_{B}/D \sim {\cal O}(10)$ at most.  As observed for rods, the isotropic-nematic ceases to exist below a critical 
ionic strength. This effect is most noticeable for disks with a continuous charge distribution along the face or rim.  For disks with a discrete hexagonal charge patterns   the window of  
stable nematic order  is somewhat larger in terms of ionic strength. The curvature of the binodals point to a reentrant phase separation phenomenon is similar to the case of rods in \fig{fig3}.     
For highly charged disks,  the suppression of nematic order is even more drastic and is borne out from the second-virial free energy \eq{b2}  using the orientation-dependent Yukawa 
potential \eq{farfield}. No stable isotropic-nematic was found in the experimentally relevant range of disk diameters $35 \lambda_{B} < D  < 200  \lambda_{B}$ and densities.  
 The lack of stability of a simple nematic state can be inferred from \fig{fig5} illustrating  the {\em effective} excluded volume $-\beta_{1}$ of a charged disk.   Although the volume depends strongly on the 
 ionic strength, its angular variation remains very weak throughout.  The effective shape of a highly charged colloidal disk  resembles that of a slightly deformed spherical object whose anisometry is 
 insufficient to generate a thermodynamically stable orientational disorder-order transition. Similar to the case of rods, it is plausable that strongly charged disks exhibit more complicated ordering mechanisms where nematicity is linked to some complex spatial
inhomogeneity of the director field as suggested by recent computer simulations \cite{jabbari2013}. These complex nematic structures call for a more local Onsager free energy  that explicitly accounts for the elastic contributions associated with spatial variations of the nematic director \cite{allenevans}.
Analogously to the charged rods, the predictive power of the Onsager-van der Waals theory  in the strong-coupling regime could be enhanced by combining an effective disk shape with a suitably chosen amplitude for the mean-field aligning potential $a_{1}(\gamma)$.

\section{Stability of liquid crystal phases with positional order}

Possible phase transitions to spatially inhomogeneous  states with smectic or columnar order can be investigated by recasting the mean-field Onsager into a functional form depending on  
the one-body density field $\rho(\br, \Omega)$ \cite{allenevans}. Within the framework of classical density-functional theory, the free energy functional needs to be minimised with respect 
to $\rho$ to yield the unique equilibrium density profile for a given chemical potential, temperature and external potential \cite{singh_pr1991}. In this work, we shall perform a simple stability 
analysis \cite{kayser_pra1978,mulder_pra1987} by assuming a weak periodic density modulation with wave-vector $\bk$ and  amplitude $\varepsilon $:
\beq
\label{mod}
\rho(\br, \Omega) = \rho_{0}f_{0}(\Omega) + \varepsilon f^{\ast}(\Omega) \cos (\bk \cdot \br),
\eeq
 superimposed onto the one-body density $\rho_{0}(\br , \Omega) = \rho_{0}f_{0}({\Omega}) $ of the spatially homogeneous phase. Beyond a particular value of  the bulk density, such a periodic 
 density perturbation will lead to  a reduction of the free energy and  the homogeneous bulk phase will become marginally unstable. The so-called bifurcation point can be found by inserting \eq{mod} 
 into the density functional and Taylor-expanding up to second order in $\varepsilon$. The resulting bifurcation condition is represented by a linear eigenvalue equation \cite{roij_pre1995}:
\beq
\label{scf}
 \int d \omb  f_{0}(\oma) \hat{\Phi} (\bk ; \oma , \omb)f^{\ast}(\omb ) = \frac{1}{\rho_{0}}f^{\ast}(\oma),
\eeq
in terms of the cosine transformed Mayer function:
\beq
 \hat{\Phi}(\bk ; \oma , \omb) = \int d \br \Phi (\br; \oma, \omb) \cos (\bk \cdot \br ).
 \eeq 
 The eigenvector $f^{\ast}(\Omega)$ probes the angular distribution in the new phase and reflects the intrinsic coupling between positional and orientational order. 
 A bifurcation to the positionally modulated state occurs at the wave vector $\bk$ that generates the smallest eigenvalue $\rho_{0} >0$ of  \eq{scf}. We reiterate that, in case of a nematic reference state,  the  ODF $f_{0}(\Omega)$ depends implicitly on $\rho_{0}$ via the self-consistency condition \eq{odfnum}.  If both the particle anisometry and density are sufficiently large, the degree of nematic order is usually very high and fluctuations in the particle orientations are strongly suppressed. Small variations in the ODF are therefore unlikely to contribute to the loss of nematic stability. In those cases it is justified to neglect the
 translation-rotation coupling and equate $f^{\ast}(\Omega) = f_{0} (\Omega)$. The bifurcation condition then takes the form of a divergence of the static structure factor $S(\bk)$:
\beq
\label{strufac}
S(\bk)^{-1} = (1 - \rho_{0} \langle \langle \hat{\Phi}(\bk ; \oma , \omb) \rangle \rangle)=0.
\eeq 
By applying the van der Waals approximation outlined in Sec. II,  $\hat{\Phi}$ can be expressed as a sum of the hard-core contribution and a part that encodes the effect of the soft potential. 
Eliminating the angular dependency of the excluded volume  $\hat{v}_{\text{excl}}$ and form factor $W$ for notational brevity, one arrives at the following expression for the Mayer kernel in 
Fourier space: 
\bea
\label{phiq}
 \hat{\Phi}(\bk ) &&= -\hat{v}_{\text{excl}}(\bk) - \hat{u}(k) \nonumber \\
&& + \frac{1}{(2 \pi)^{3}}  \int d \bq  \hat{u}(q)  W(\bq  )W (-\bq ) \hat{v}_{\text{excl}}(\bk - \bq ).
\eea
The Fourier integral presents a non-trivial mode-coupling term that convolutes the imposed density wave with the modes describing the distance-dependence of the soft interactions. 
The solution of \eq{strufac} (or \eq{scf}) for particles with full orientational degrees of freedom poses a substantial technical task and we shall simplify matters by considering the 
more tractable case of {\em parallel} cylinders.  Let us equate the particle frame to the lab frame $\{ \bx, \by, \bz \} $ with the cylinder normals pointing along $\bz$.  The excluded 
volume of two parallel cylinders is again a cylinder with volume $2 \pi LD^{2}$.  In Fourier space the excluded volume takes the following form:
\beq
\hat{v}_{\text{excl}}(\bq) = 2 \pi L D^{2} j_{0} (L \bq \cdot \bz) \frac{J_{1}(\sqrt{(D\bq \cdot \bx)^{2}+(D \bq \cdot \by)^{2}})}{ \frac{1}{2}\sqrt{(D\bq \cdot \bx)^{2}+(D \bq \cdot \by)^{2}}.
}
\eeq
 Due to the parallel orientation it is no longer possible to take the limit of infinite particle anisometry since the excluded volume vanishes in both limits (similar to setting  
 $\gamma  =0$ in \eq{vexcluded}). Therefore we shall consider the case $D/L \ll x$ (rods) and $L/D \ll x$ (disks) with $x$ a small but finite number and use the volume fraction 
 $\phi = (\pi /4) LD^{2}\rho_{0}$ as a convenient measure for the particle concentration. 

We may probe instabilities pertaining to smectic order by identifying $\bk =k_{S} \{ 0,0, 1\}$,  a one-dimensional periodic modulation along the nematic director. Hexagonal columnar 
order can be parametrized by a linear superposition of three modulations with wave-vectors $\bk_{1}= k_{C} \{  0,1,0  \}$, $\bk_{2} =k_{C} \{ \frac{\sqrt{3}}{2}, \frac{1}{2}, 0 \}  $, 
and $\bk_{3} =k_{C} \{ -\frac{\sqrt{3}}{2}, \frac{1}{2}, 0 \}  $ describing a two-dimensional triangular lattice perpendicular to the director.

\begin{figure}
\begin{center}
\includegraphics[clip=,width= 0.8\columnwidth]{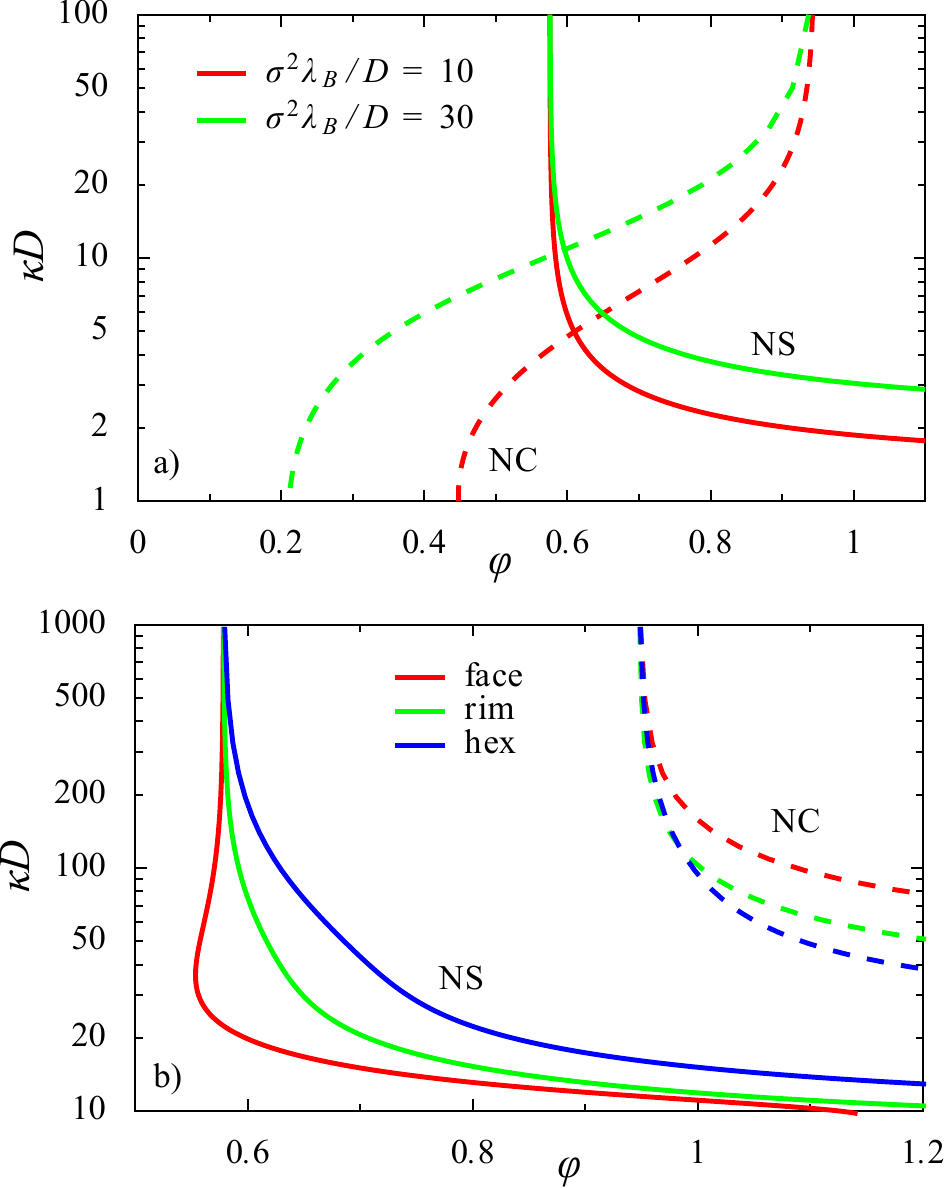}
\caption{ \label{fig6} (a) Variation of the nematic-smectic (NS) and nematic-columnar (NC) bifurcation density with ionic strength $\kappa D$ for a) parallel charged rods with 
aspect ratio $L/D=50$ and (b) parallel disks with surface charge patterns indicated in Fig. 1 ($D/L=10$, $Z^{2}\lambda_{B}/D = 10$). Solid curves indicate  Nematic-Smectic (NS) 
bifurcations, dotted curves Nematic-Columnar (NC) instabilities. }
\end{center}
\end{figure}

The results  in \fig{fig6} reveal a marked stabilization of columnar with respect to smectic order for rodlike cylinders at low ionic strength. This outcome is in accordance 
with previous numerical results for Yukawa rods in a strong external aligning field \cite{wensink2007}. Needless to say that the transition values are merely qualitative and that the 
volume fractions can be brought down to more realistic values, for instance, by using an effective second-virial theory based on a resummation of higher virial coefficient  (e.g. using Parsons'  theory \cite{parsons1979}). 
For hard parallel disks, the nematic-smectic always pre-empts the nematic-columnar one irrespective of the aspect ratio $x$.  This implies that the parallel approximation fares rather badly 
for hard discotic systems  which are known to form columnar phases only \cite{Veerman,vanderkooij_jpcb1998}.  Nevertheless some general trends for can be gleaned from \fig{fig6}b such as an 
apparently stabilization of smectic order for uniformly charged disks in the low screening regime. The prevalence of smectic order  has been recently reported in weakly screened discotic 
systems \cite{kleshchanok_jacs2012}. As for the other charge patterns, the observation from \fig{fig6} that both smectic and columnar-type order are destabilized upon reducing the screening 
could hint at more complicated instability mechanisms prevailing in the low screening region, such as those pertaining to crystalline order where both longitudinal and transverse density 
modulations compete with spatial inhomogeneities in the director field \cite{jabbari2013}.     

We wish to emphasize that the approach outlined above is generic in that it provides a simple  route to gauge the effect of soft interactions on the stability of positionally ordered 
liquid crystals.  It can be applied to a vast range of model systems with various segment potentials (provided integrable) and form factors. Instabilities from nematic to other 
liquid crystals symmetries or three-dimensional crystals (e.g. fcc or bcc) can easily be included by  adapting the $\bk$ vectors to the desired Bravais lattice. 
In order to describe fully crystalline states we may exploit the fact that particles are strongly localized around their lattice site to construct an appropriate density functional 
representation for the excess Helmholtz free energy. In the following, we shall briefly sketch the approach outlined in Refs. \cite{xulekkerkerker,archer_pre2005}. The central 
assumption is that  the density profile of the solid consists of Gaussian peaks centred on a predefined lattice vector  $\{ {\bf R}_{i} \}$ factorized with the orientational probability 
(ODF) $f(\Omega)$.  If we assume a spatially homogeneous director field, the one-body density can be written as: 
\beq 
\label{pargauss}
\rho (\br, \Omega) = f(\Omega) \sum_{i=1}^{N} G(\br - {\bf R}_{i}),
\eeq
with
\beq
\label{gauss}
G(\br - {\bf R}_{i}) = \left (\frac{\alpha}{\pi} \right )^{3/2} \exp [ - \alpha (\br - {\bf R}_{i})^{2} ],
\eeq
where $\alpha$ is a parameter which describes how localised the
particles are around each lattice site. Assuming the proportion of lattice defects to be negligible, each lattice site should contain only one particle as reflected in the 
normalisation of \eq{gauss}. The excess free energy of the system can be expressed in terms of the following Fourier integral:
\beq
\label{fexgauss}
F_{\text{ex}} = -  \frac{k_{B}T}{2} \sum_{i \neq j} \frac{1}{(2 \pi)^{3}} \int d \bk e^{i \bk \cdot {\bf R}_{ij}}   \hat{G}(k)^{2} \langle \langle  \hat{\Phi}(\bk ; \Omega, \Omega^{\prime}) \rangle \rangle, 
\eeq
which is composed of the FT of the orientation-dependent  Mayer kernel \eq{phiq}, and the Gaussian weight $\hat{G}(k) =  \exp (-k^{2} /4\alpha )$.  In general,  the radially symmetric 
form \eq{gauss}  is justified only if particles are strongly localized around their lattice points ($\alpha \gg 1$) so that the density peaks are not affected by the symmetry of 
the underlying lattice.  The total free energy is obtained by combining the excess free energy with the ideal free energy associated with the Gaussian parameterization: 
\beq
F_{\text{id}} = N k_{B}T \left \{ \frac{3}{2} \ln \left (\frac{{\mathcal V}^{2} \alpha}{\pi} \right )-  \frac{5}{2}  + \langle  \ln f (\Omega) \rangle \right \}.
\eeq
Next, the free energy must be minimized with respect to the localization parameter $\alpha$,  the set of relevant lattice constants corresponding to the imposed lattice 
symmetry \cite{baus_mp1983} and $f(\Omega)$. This simple variational scheme  allows one to compare the stability of various crystal symmetries  as a function of density 
and interaction range and strength. In addition, due to the translation-orientation coupling via $f$, both aligned and rotationally disordered plastic crystal states can be included.  
Phase transitions between fluid and crystal phases can be probed by equating the pressure and chemical potential emerging from the Gaussian free energy with those of the fluid phases, \eq{presmu}.  

\section{Concluding remarks}

We have proposed a generalized Onsager theory for strongly non-spherical colloidal particles with an intrinsic patchiness in the interaction potential. The theory supplements the  
second-virial reference free energy for the hard-core interaction with a first-order perturbative (van der Waals) term, which captures the directional soft interactions between the 
rods or the disks. As such, the theory interpolates between the low density regime, where the second-virial approximation holds, and the high density regime where the mean-field 
approach is accurate.  We have aimed at formulating a generic framework that should be applicable to a wide range of particle shapes, ranging from elongated rods to flat, sheet-like 
disks with an arbitrary spatial organization of interactions sites distributed along the colloid surface. By recasting the mean-field contribution in terms a  Fourier series, the excess 
free energy naturally factorizes into three main contributions: the site-site interaction potential, the shape of the colloidal hard-core, and a form factor associated with the spatial 
arrangement of the interaction site residing on each particle. 

As a test case, we have applied our theory to  investigate orientation disorder-order transitions in fluids of charged rods and disks with a uniform, localized or  discretized surface 
charge  pattern. The results for the isotropic-nematic phase diagram and the instability analysis of transverse and longitudinal freezing of a nematic fluid in  the 
high-density regime, reveal 
a picture that is consistent with results from more elaborate Poisson-Boltzmann approaches and particle simulation. This lends credence to our theory as a practical tool to assess the 
influence of soft patchy interactions on the liquid crystal phase diagram of non-isometric colloids.  Although  the focus of this study is on the liquid crystal fluid phases that emerge 
at relatively low particle density, the stability of spatially ordered liquid crystals at higher particle concentration can also be scrutinized using a simple bifurcation analysis while 
fully crystalline phases can be expediently accounted for using a Gaussian parameterization for the one-body density often used in density functional theories of freezing.

We remark that the present theory is amenable to various extensions towards more complicated systems. Colloidal dispersions composed of non-spherical particles are rarely monodisperse 
but are often characterized by a  continuous spread in particle sizes.  The polydisperse nature of the colloid shape and/or  the amplitude of the soft interactions can be incorporated in 
a straightforward manner \cite{baus_jpcm2002,wensink_jcp2003}.  Bio-colloids such as stiff viral rods \cite{grelet_prl2003} and DNA  are commonly characterized  by an intrinsic helical 
patchiness which has profound implications on the mesostructure in bulk and confinement \cite{leforestier_crc2008}.  The present  theory could be be extended to relate the mesoscopic 
chirality of twisted nematics to the intrinsic helical form factor of the colloid  \cite{wensink_jpcm2011}.

Last but not least, similar to systems of spherical subunits  \cite{malijevsky_pre2005,roth_mp2011},  more accurate reference free energies could be employed which should give a more 
reliable account of correlations in systems of less anisometric colloids (dumbbells, thick platelets, polyhedra), which routinely form highly ordered (liquid) crystals at high particle 
volume fraction \cite{vega_dumbbell_jcp1992,glotzer_polyhedra2012}.

\section*{Appendix: excluded volume of strongly anisometric cylinders}

\begin{figure}
\begin{center}
\includegraphics[clip=,width= 0.9\columnwidth]{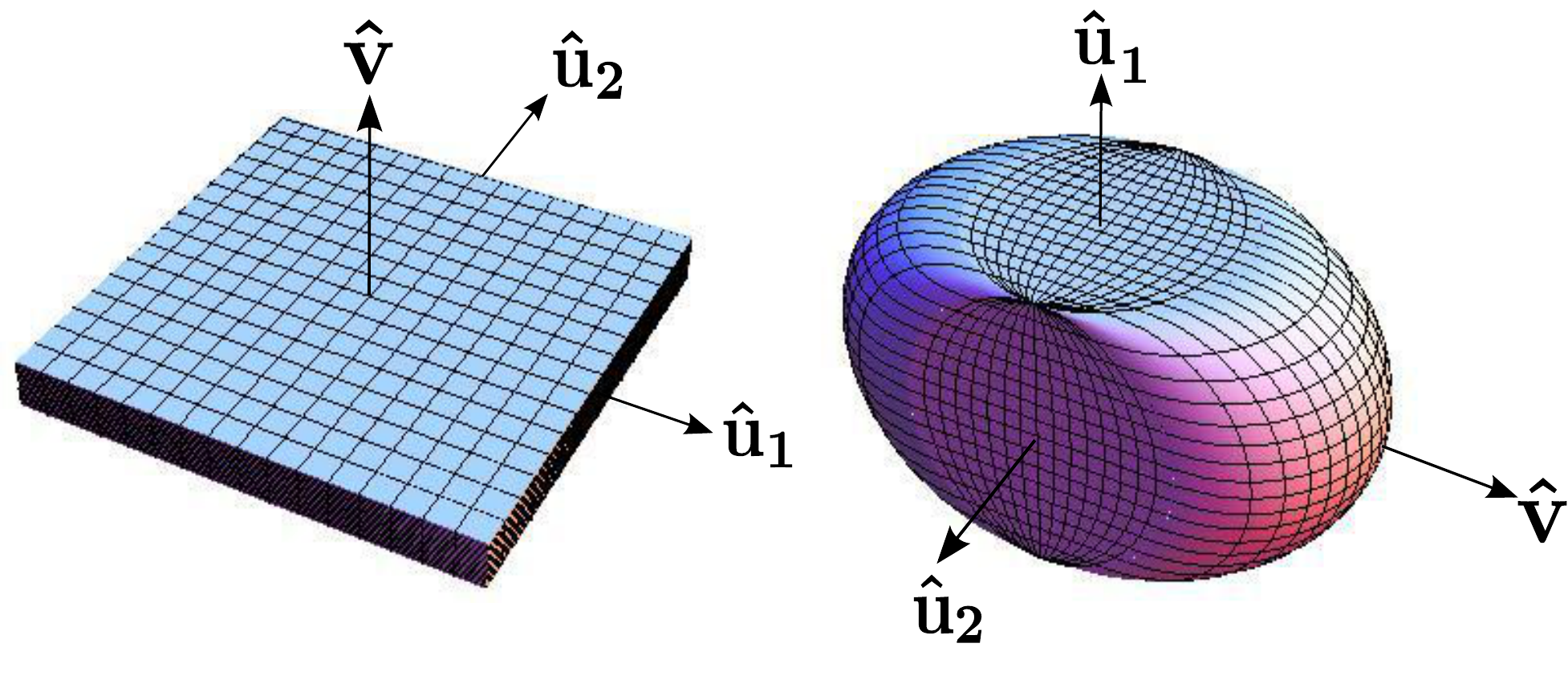}
\caption{ \label{fig7} The excluded-volume manifold of two infinitely slender cylindrical rods ($L/D \rightarrow \infty$) is a parallelepiped spanned by the particle-based coordinate 
frame $ \{ \bhua, \bhub, \bv \} $ (left figure) whereas that of two infinitely thin disks ($ L/D \downarrow 0$) is represented by a sphero-cuboid (right figure).  
Both manifolds correspond to the case where the cylinders are perpendicular to each other  ($\bhua \perp \bhub$).}
\end{center}
\end{figure}

In this Appendix, we derive expressions for the Fourier Transform (FT) of the excluded volume manifold of two infinitely slender rods and disks,  featured in \eq{a1} of the main text. 
The excluded volume of two hard cylinders at fixed angle $\gamma$ is a {\em parallelepiped}  which can be parameterized by switching from the laboratory frame to a particle frame 
spanned by the normal orientational unit vectors $\bhu_{\alpha}$ of the cylinder pair. Let us define the additional unit vectors:
\bea
\bv   | \sin \gamma | & = &  \bhua \times \bhub    \nonumber \\
\bw _{\alpha} & = &   \bhu _ {\alpha} \times \bv    \hspace{0.5cm} (\alpha =1,2) \label{unit},
\eea
so that $ \{ \bhu_{\alpha} , \bv , \bw_{\alpha} \}$  are two orthonormal basis sets in 3D. The centre-of-mass distance vector can be uniquely decomposed in terms of these basis vectors:
\beq
\br  =  ( \br \cdot \bhu_{\alpha} )  \bhu_{\alpha}  +  ( \br \cdot \bv )  \bv  +  ( \br \cdot \bw_{\alpha} )  \bw_{\alpha}.
\eeq
The leading order contribution to the excluded-volume body  is of
${\cal O}(L^{2}D)$ and stems from the overlap of the cylindrical parts of the cylinders. This resulting parallelepiped can be parameterized as follows:
\beq
\br _{cc} =  \frac{L}{2} t_1 \bhua  + \frac{L}{2} t_2 \bhub   + D t_3 \bv,  
\eeq
with  $-1 \le  t_i \le 1 $ for $i=1,2,3$. The Jacobian associated with the coordinate transformation is $J_{cc}= \frac{1}{4} L^{2} D | \sin \gamma | $. The FT of the parallelepiped is thus given by:
\bea
\label{cotrafo}
\hat{v}_{\text{excl}} (\oma , \omb ) & = & \int d \br_{cc} e^{i \bq \cdot \br_{cc}} \nonumber \\
& =& J_{cc} \prod_{i<3} \int_{-1}^{1} dt_i \cos (\bq \cdot \br_{CC}) \nonumber \\
& = & v_{0} | \sin \gamma | {\mathcal F} ( \bq ; \oma , \omb),
\eea
where $v_{0} = 2L^{2}D$. Using that $\int_{-1}^{1} dx \cos (ax + b) = 2j_{0}(x) \cos b $ one obtains for strongly elongated cylinders
(needles):
\beq
\label{frods}
{\mathcal F} (\bq ; \oma, \omb) = j_{0} \left( \frac{L}{2} \bq \cdot \bhua   \right) j_{0} \left( \frac{L}{2} \bq \cdot \bhub  \right) j_{0} ( D \bq \cdot \bv ), 
\eeq
in terms of the spherical Bessel function $j_{0} (x) = \sin x/x$. A similar procedure can be carried out for disk-shaped cylinders. Two infinitely flat cylinders overlap if the separation $\br$ of 
their centre-of-mass  is in a sphero-cuboid (see \fig{fig7}) which can be parameterized as follows: 
\beq
\label{cuboid}
\br_{cc} = - \frac{D}{2} t_1 \bw_{1} - \frac{D}{2} t_2 \bw_{2} +\frac{D}{2} t_{3} [(1-t_{1}^{2})^{1/2} + (1- t_{2}^{2})^{1/2} ] \bv,
\eeq
with  $-1 \le  t_i \le 1 $ for $i=1,2,3$. The Jacobian associated with the transformation from the lab to the particle frame is 
$d \br _{cc}  = J_{cc} dt_{1}d_{2}dt_{3} $ with $J_{cc} = \frac{D^{3}}{8} | \sin \gamma | [(1-t_{1}^{2})^{1/2} + (1- t_{2}^{2})^{1/2} ] $. Similar to the case of rods the FT of the excluded volume 
figure is  cast into a cosine transform according to \eq{cotrafo} substituting $v_{0} = \pi D^{3}/2$ for disks. The shape function ${\mathcal F}$, however, requires a bit more effort in this case. 
First, the integration over $t_{3}$ can be carried out straightforwardly using the relation involving the spherical Bessel function mentioned above \eq{frods}. This yields:
\bea
{\mathcal F} &=&  \frac{1}{\pi D \bq \cdot \bv } \int_{-1}^{1}dt_{1} \int_{-1}^{1} dt_{2}  \cos ( \frac{D}{2} t_1 \bq \cdot \bw_{1} +  \frac{D}{2} t_2 \bq \cdot \bw_{2}   ) \nonumber \\
&& \times \sin \{ [ (1-t_{1}^{2})^{1/2}  + (1 - t_{2}^{2})^{1/2} ] \frac{D}{2} \bq \cdot \bv \}. 
\eea
The double integral can be split into single integrals using standard trigonometric manipulations. Rearranging terms gives the final expression for infinitely flat  discotic cylinders (disks): 
\bea
{\mathcal F}(\bq ; \oma , \omb) &=&  \frac{1}{\pi D \bq \cdot \bv } (A_{1} B_{2}  + A_{2} B_{1}), 
\eea
in terms of the orientation-dependent functions:
\bea
\label{ab}
A_{\alpha} &=& \int_{-1}^{1} dt \cos (\frac{D}{2}t \bq \cdot \bw_{\alpha}) \cos (\frac{D}{2} (1-t^{2})^{1/2} \bq \cdot \bv) \nonumber \\
B_{\alpha} &=& \int_{-1}^{1} dt \cos (\frac{D}{2}t \bq \cdot \bw_{\alpha}) \sin (\frac{D}{2} (1-t^{2})^{1/2} \bq \cdot \bv).   
\eea
The last integral can be solved in closed form by substituting $t = \cos \theta $ and invoking Catalan's integral representation of Bessel functions \cite{bateman1954tables}:
\beq
  J_{0} ( \sqrt{ \beta^{2} - \alpha^{2}} ) = \frac{1}{\pi}\int_{0}^{\pi} d \theta e^{\alpha \cos \theta} \cos ( \beta \sin \theta ),
\eeq
with $J_{n}(x)$ a Bessel function of the first kind. With this, the solution of \eq{ab} can be found by taking the partial derivative to $\alpha$ on both sides and rearranging terms:
\beq
B_{\alpha} = \frac{\pi}{2}  (\bq \cdot \bv ) J_{1}(\tilde{q}_{\alpha})/(\tilde{q}_{\alpha}),
\eeq
where $ \tilde{q}_{\alpha} =  \left [( \frac{D}{2} \bq \cdot \bw_{\alpha}  )^{2} + ( \frac{D}{2} \bq \cdot \bv  )^{2} \right ] ^{1/2}$.  Despite the similarity between $A$ and $B$ there is no 
closed analytical expression available for $A$  but the one-dimensional integral is readily evaluated using standard numerical integration routines.

\bibliographystyle{apsrev}


\end{document}